\documentclass[letter]{aa} 

\usepackage{amsmath}
\usepackage{graphicx}
\usepackage{graphics}
\usepackage{txfonts}
\usepackage{natbib}
\usepackage{xcolor}
\usepackage{url}

\bibpunct{(}{)}{;}{a}{}{,}

\def\teff{\ifmmode T_{\rm eff} \else $T_{\mathrm{eff}}$\fi}

\def\ltsima{$\buildrel<\over\sim$}
\def\lsim{\lower.5ex\hbox{\ltsima}}

\newcommand{\hii}{H~{\sc ii}}
\newcommand{\ha}{\ifmmode {\rm H}\alpha \else H$\alpha$\fi}
\newcommand{\hb}{\ifmmode {\rm H}\beta \else H$\beta$\fi}
\newcommand{\hg}{\ifmmode {\rm H}\gamma \else H$\gamma$\fi}
\newcommand{\lya}{\ifmmode {\rm Ly}\alpha \else Ly$\alpha$\fi}
\newcommand{\hei}{He~{\sc i}}

\newcommand{\ebv}{\ifmmode E_{\rm B-V} \else $E_{\rm B-V}$\fi}
\newcommand{\av}{\ifmmode A_{\rm V} \else $A_{\rm V}$\fi}
\def\micron{$\mu$m}

\def\msun{\ifmmode M_{\odot} \else M$_{\odot}$\fi}
\def\msunyr{\ifmmode M_{\odot} {\rm yr}^{-1} \else M$_{\odot}$ yr$^{-1}$\fi}
\def\zsun{\ifmmode Z_{\odot} \else Z$_{\odot}$\fi}

\def\lsun{\ifmmode L_{\odot} \else L$_{\odot}$\fi}

\def\mup{\ifmmode M_{\rm up} \else M$_{\rm up}$\fi}
\def\mlow{\ifmmode M_{\rm low} \else M$_{\rm low}$\fi}

\def\aap{A\&A}

\def\apjl{ApJL}
\def\apjs{ApJS}

\newcommand{\oh}{\ifmmode 12 + \log({\rm O/H}) \else$12 + \log({\rm
O/H})$\fi}

\newcommand{\oii}{[O~{\sc ii}]}
\newcommand{\oiii}{[O~{\sc iii}]}
\newcommand{\neiii}{[Ne~{\sc iii}]}

\def\Oii{[O~{\sc ii}] $\lambda$3727}
\def\Oiii{[O~{\sc iii}] $\lambda\lambda$4959,5007}
\def\Oiiib{[O~{\sc iii}]$\lambda 5007$}
\def\Oiiit{[O~{\sc iii}]$\lambda 4363$}
\newcommand{\Neiii}{[Ne~{\sc iii}] $\lambda$3869}

\def\flyf{\ifmmode f_{\rm Lyf} \else $f_{\rm Lyf}$\fi}
\def\pz{\ifmmode P(z) \else $P(z)$\fi}
\def\ki2{\ifmmode \chi^2 \else $\chi^2$\fi}
\def\zphot{\ifmmode z_{\rm phot} \else $z_{\rm phot}$\fi}

\newcommand{\xphot}{\ifmmode x_\gamma \else $v_\gamma$\fi}
\newcommand{\xobs}{\ifmmode x_{\rm obs} \else $x_{\rm obs}$\fi}
\newcommand{\xcmf}{\ifmmode x_{\rm CMF} \else $x_{\rm CMF}$\fi}
\newcommand{\vexp}{\ifmmode V_{\rm exp} \else $V_{\rm exp}$\fi}
\newcommand{\vmax}{\ifmmode V_{\rm max} \else $V_{\rm max}$\fi}
\newcommand{\nh}{\ifmmode N_{\rm HI} \else $N_{\rm HI}$\fi}
\newcommand{\dv}{\ifmmode \Delta v({\rm em-abs}) \else $\Delta v({\rm em}-{\rm abs})$\fi}

\def\fesc{\ifmmode f_{\rm esc} \else $f_{\rm esc}$\fi}
\def\fescrel{\ifmmode f_{\rm esc,rel} \else $f_{\rm esc,rel}$\fi}

\def\frellya{\ifmmode f^{\rm rel}_{\rm{Ly}\alpha} \else $f^{\rm rel}_{\rm{Ly}\alpha}$\fi}

\def\hii{H{\sc ii}}

\newcommand{\mstar}{\ifmmode M_\star \else $M_\star$\fi}
\newcommand{\muv}{\ifmmode M_{1500} \else $M_{1500}$\fi}
\newcommand{\auv}{\ifmmode A_{\rm UV} \else $A_{\rm UV}$\fi}
\newcommand{\luv}{\ifmmode L_{\rm UV} \else $L_{\rm UV}$\fi}
\newcommand{\lir}{\ifmmode L_{\rm IR} \else $L_{\rm IR}$\fi}
\newcommand{\lbol}{\ifmmode L_{\rm bol} \else $L_{\rm bol}$\fi}
\newcommand{\liruv}{\ifmmode L_{\rm IR+UV} \else $L_{\rm IR+UV}$\fi}
\newcommand{\liroveruv}{\ifmmode L_{\rm IR}/L_{\rm UV} \else $L_{\rm IR}/L_{\rm UV}$\fi}
\newcommand{\nlyc}{\ifmmode N_{\rm Lyc} \else $N_{\rm Lyc} $\fi}
\newcommand{\rholyc}{\ifmmode \rho_{\rm Lyc} \else $\rho_{\rm Lyc} $\fi}
\newcommand{\chion}{\ifmmode \xi_{\rm ion} \else $\xi_{\rm ion}$\fi}
\newcommand{\chioncorr}{\ifmmode \xi_{\rm ion}^0 \else $\xi_{\rm ion}^0$\fi}

\begin{document}

\title{First look with JWST spectroscopy: $z \sim 8$ galaxies resemble local analogues}
\subtitle{}
\author{D. Schaerer\inst{1,2}, 
R. Marques-Chaves\inst{1}, 
L. Barrufet\inst{1}, 
P. Oesch\inst{1,3}, 
Y. I. Izotov$^{4}$,
R. Naidu$^{5}$, 
N. G. Guseva$^{4}$,
G. Brammer\inst{3}
}
  \institute{Observatoire de Gen\`eve, Universit\'e de Gen\`eve, Chemin Pegasi 51, 1290 Versoix, Switzerland
         \and
CNRS, IRAP, 14 Avenue E. Belin, 31400 Toulouse, France
	\and
Cosmic Dawn Center (DAWN), Niels Bohr Institute, University of Copenhagen, Jagtvej 128, K\o benhavn N, DK-2200, Denmark
\and
Center for Astrophysics, Harvard \& Smithsonian, 60 Garden Street, Cambridge, MA 02138, USA
\and
Bogolyubov Institute for Theoretical Physics,
National Academy of Sciences of Ukraine, 14-b Metrolohichna str., Kyiv,
03143, Ukraine
         }

\authorrunning{D.\ Schaerer et al.}
\titlerunning{}

\date{Accepted for publication in A\&A Letters}


\abstract{Deep images and near-IR spectra of galaxies in the field of the lensing cluster SMACS J0723.3-7327 were recently
taken in the Early Release Observations program of JWST.
Among these, two NIRSpec spectra of galaxies at $z=7.7$ and one at $z=8.5$ were obtained, revealing for the first time
rest-frame optical emission line spectra of galaxies in the epoch of reionization, including the detection of the important \Oiiit\ auroral line (see JWST PR 2022-035).
We present an analysis of the emission line properties of these galaxies, finding that these galaxies have a high 
excitation (as indicated by high ratios of \Oiiib/\Oii, \Neiii/\Oii), strong \Oiiit, high equivalent widths,
and other properties which are typical of low-metallicity star-forming galaxies. 
Using the direct method we determine oxygen abundances of $\oh \approx 7.9$ in two $z=7.7$ galaxies, and a lower metallicity of $\oh \approx 7.4-7.5$ ($\sim 5$ \% solar) in the $z=8.5$ galaxy using different strong line methods. More accurate metallicity determinations will require better data.
With stellar masses estimated from SED fits, we find that the three galaxies lie close to or below the $z \sim 2$ mass-metallicity relation.  
Overall, these first galaxy spectra at $z \sim 8$ show a strong resemblance of the emission lines properties
of galaxies in the epoch of reionization with those of relatively rare local analogues previously studied from the SDSS.
Clearly, the first JWST observations demonstrate already the incredible power of spectroscopy to reveal properties of galaxies in  the early Universe.
}


 \maketitle

\section{Introduction}
\label{s_intro}
Optical emission line spectroscopy has long provided important insights on the physical composition, 
properties of the interstellar medium (ISM), and the nature of the ionizing power of galaxies, yielding thus
key information to understand many key aspects of galaxy evolution \citep[see review of][]{Kewley2019Understanding-G}.
The well-known emission lines of H, He, O, N, S, Ne, and other elements detected in optical galaxy spectra, emitted in the ionized ISM (\hii\ regions primarily) have been detected in nearly one million of galaxy spectra, out to redshifts of $z \sim 0.5-1$ with the Sloan Digital Sky Survey \citep[SDSS,][]{Ahumada2020The-16th-Data-R}. 
Ground-based near-IR spectroscopy has recently pushed these
limits to $z \sim 1.5-3$, where $\sim 1500$  measurements of the strongest rest-optical lines have
been possible, e.g.\ with the MOSDEF survey \citep{Kriek2015THE-MOSFIRE-DEE}, revealing thus ISM properties
at cosmic noon \citep[e.g.][]{Forster-Schreiber2020Star-Forming-Ga}.

The recent launch of the JWST and the spectroscopic capabilities of its multi-object near-infrared spectrograph (NIRSpec)
in the opens now a completely new window into the early Universe, where, for the first time, 
all the ``classical'' optical diagnostics developed at low-$z$ can be used to study the properties of 
galaxies over a wide redshift range, from $z \sim 3$ out to the epoch of reionization ($z > 6.5$).

The first public NIRSpec observations, part of the Early Release Observations (ERO) of JWST, have covered the 
SMACS J0723.3-7327 galaxy cluster, providing $1.8-5.2$ \micron\ spectra of 35 objects in 
the field. Among those, three galaxies in the epoch of reionization were observed, showing spectacular,
rich rest-frame optical emission line spectra of objects at $z=7.7$ and $z=8.5$
(see JWST Press release 2022-035\footnote{\url{https://webbtelescope.org/contents/news-releases/2022/news-2022-035}}).
%
We here report the first determination of the metallicity (O/H) of these galaxies, a detailed 
analysis of their emission line properties, and a comparison with observed properties of low-$z$ 
emission line galaxies.
At the time of revision, the JWST targets have been the object of other studies analysing their emission
line spectra and derived properties \citep[see][]{Arellano-Cordova2022A-First-Look-at,Brinchmann2022High-z-galaxies,Carnall2022A-first-look-at,Curti2022The-chemical-en,Katz2022First-Insights-,Rhoads2022Finding-Peas-in,Taylor2022Metallicities-o,Trump2022The-Physical-Co} .

\section{Observations}
\label{s_obs}

\begin{figure*}
  \centering
  \includegraphics[width=0.98\textwidth]{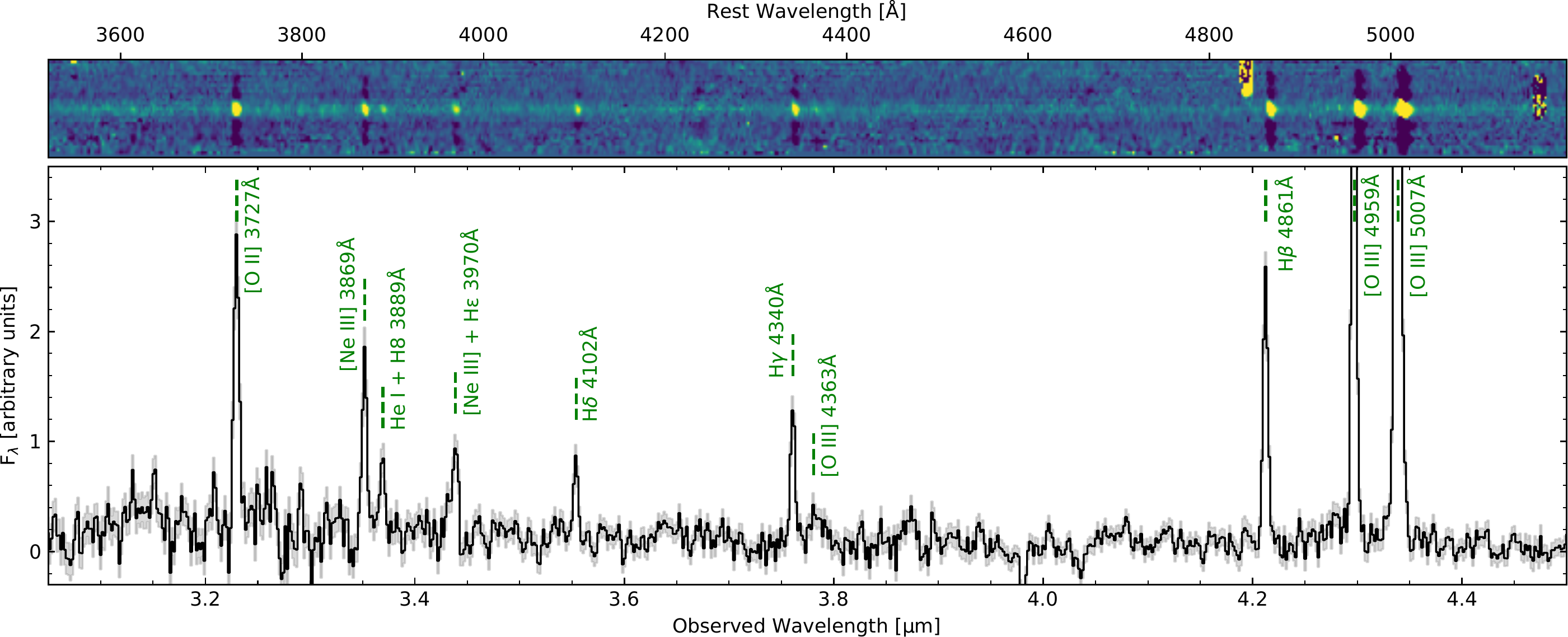}
  \caption{2D (top) and 1D (bottom) NIRSpec/JWST spectrum of 06355 at $z=7.664$ (black)  and 1$\sigma$ uncertainty (grey). Vertical dashed lines mark the position of well-detected nebular emission lines. The continuum emission is also detected as seen in the 2D spectrum. X-axis in the bottom and top panels refer to the observed ($\mu$m) and rest-frame wavelengths (\AA), respectively. 
  }
  \label{fig_spec}
\end{figure*}

\begin{figure}[htb]
  \centering
  \includegraphics[width=0.45\textwidth]{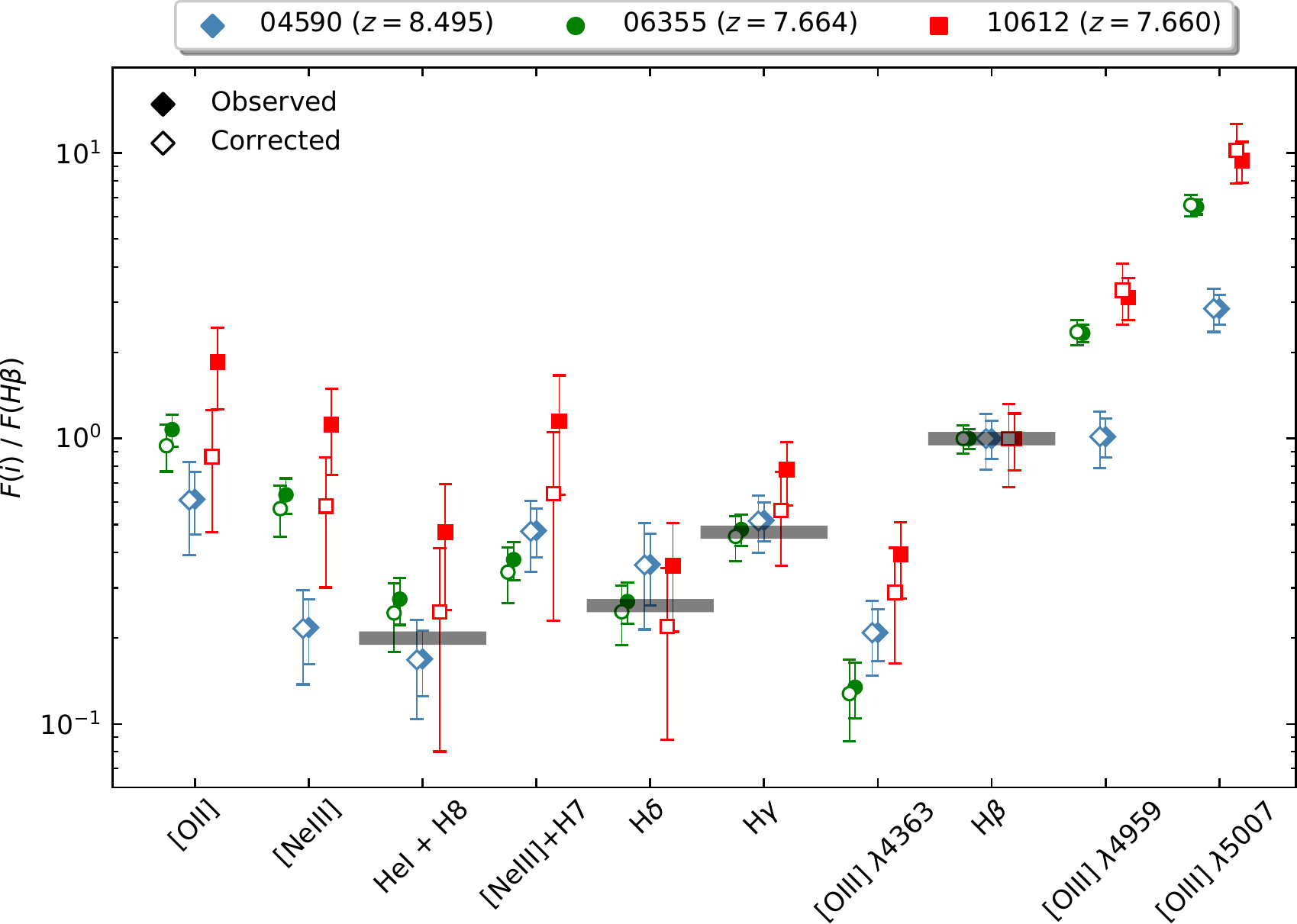}
  \caption{Relative line intensities of the measured lines with respect to \hb. Full and open symbols show the ratios prior to and after applying our flux correction (see text).}
  \label{fig_fluxes}
\end{figure}

\subsection{JWST NIRSpec observations}

Rest-frame UV and optical spectra were obtained on 30 June 2022 using NIRSpec with the micro-shutter assembly (MSA). Observations consist of two different pointings (s007 and s008), each of them using two grating/filter combinations: G235M/F170LP and G395M/F290LP. The total exposure time is 2$\times 8754$ seconds for each grating/filter. This provides a spectral resolution $R \simeq 1000$ and a continuous spectral coverage of $\simeq 1.75 - 5.20\mu$m. Fully calibrated spectra (calibration level 3) were retrieved from the Mikulski Archive for Space Telescopes\footnote{\url{https://mast.stsci.edu/}}, which were previously processed with the JWST Science Calibration Pipeline ({\sc cal\_ver}: 1.5.3 and {\sc crds\_ctx}:  jwst-0916.pmap). For each source, two individual 1D spectra (s007 and s008) are combined using the average flux, and after masking spectral regions affected by cosmic rays and other artifacts, 
after inspection of the 2D spectra. For 04590 we exclude data above $\lambda_{\rm obs} \geq 4.5 \mu$m (i.e., after the detector gap) for the observation s008 as the 2D spectrum in this region shows several artifacts that might compromise the flux calibration, including a considerable flux gradient along the spatial direction. 
An example of the combined spectrum of 06355 at $z \simeq 7.7$ is shown in Fig.~\ref{fig_spec}. 

The spectra of the three sources show nebular emission lines, many of them detected with high significance ($>3\sigma$). These include Balmer lines (H$\beta$, H$\gamma$, H$\delta$), [O~{\sc ii}] and [O~{\sc iii}], [Ne~{\sc iii}], and He~{\sc i}. In particular, the auroral \Oiiit\ line -- key for accurate determinations of the O/H abundance using the direct method -- is detected in all galaxies with a significance of $3.9-5.7 \sigma$.
Gaussian profiles are fitted to each line using the Python nonlinear least-squares function {\sc curve-fit}, and assuming a constant level for the continuum (in $f_\lambda$). We derive the redshifts for these sources using the brightest lines, namely H$\gamma$, H$\beta$, and \Oiii\ lines (Table~\ref{tab_1}). The continuum is clearly detected in the two $z\simeq 7.7$ sources (10612 and 06355, see Fig.\ref{fig_spec}) allowing the determination of the equivalent widths of the lines (Table~\ref{tab_1}). However, these do not include any correction for slit-losses and other possible effects due to different morphologies of the continuum emission and nebular lines. 

After careful inspection of the flux measurements we noticed that some line ratios have  nonphysical values, suggesting systematics on the flux calibration and throughput in the current JWST pipeline. For example, the observed 
Balmer line ratios are found to be larger than the intrinsic H$\gamma$/H$\beta$ and H$\delta$/H$\beta$ ratios assuming case B recombination. Before more accurate calibration reference files are available, we overcame this issue by applying an ad hoc correction to the flux calibration. More specifically, we fit a power law ($\propto \lambda^{\alpha}$) to i) the intrinsic H$\gamma$/H$\beta = 0.47$, H$\delta$/H$\beta = 0.26$ (case B),  and (He~{\sc i}+H8)/H$\beta = 0.2$ and to ii) the observed line ratios.
The correction factor as a function of wavelength is thus given by the division between i) and ii) for each source, and it is applied to the full spectral range covered by the G235M/F170LP grating/filter configuration. 
This empirical correction results in a maximum increase of $\sim 20-50$\% (depending on galaxy) for the O32=[O~{\sc iii}]~$\lambda$5007/[O~{\sc ii}]~$\lambda$3727 ratio, and smaller corrections for other line ratios. By doing this, we are also applying a first-order correction to the Balmer decrement, thus losing the information on the dust attenuation. 
Given the uncertainties on the flux calibration and on our empirical correction, we conservatively add $20$\% of uncertainties on our flux measurements,
which should also account for other sources of uncertainty, such as those arising from the estimation of the continuum level. This depends on the different assumptions (e.g., size of the spectral windows used in the fit, constant level or straight line, noise fluctuations, etc.) and may affect the flux measurements of different lines.
The measured line ratios, prior to and after correction, are shown in Fig.\ \ref{fig_fluxes}.
Where possible, our analysis will focus mostly on lines ratios between nearby lines (see Section~\ref{s_discuss}), thus minimizing possible issues on the flux calibration, and dust attenuation.

It must be noted that presently the data reduction is still in a preliminary stage, which uses in particular pre-launch calibration files \citep{Rigby2022Characterizatio}. To tackle with these limitations, different groups have used different approaches, including observations of a standard star, different extractions of the 2D spectra, different combinations of the 2 visits and other corrections  \citep[cf.][]{Curti2022The-chemical-en,Rhoads2022Finding-Peas-in,Trump2022The-Physical-Co}.
A comparison shows differences in several line ratios, by amounts which are sometimes larger than the quoted uncertainties.
To illustrate this, we subsequently compare our measurements with those from \cite{Curti2022The-chemical-en} who have improved the calibration using a standard star.
For other comparisons see also \cite{Taylor2022Metallicities-o} and \cite{Brinchmann2022High-z-galaxies}.

\subsection{JWST NIRCam observations}

We make use of a single NIRCam pointing in the six wide filters F090W, F150W, F200W, F277W, F356W, and F444W, with a uniform exposure time of 2.1hr in each, and shallower NIRISS imaging in F115W.
We obtained the calibrated and distortion-corrected NIRCam and NIRISS images from the publicly available reduction by \cite{Brammer2022Images-and-cata}.
The images were processed with the standard JWST pipeline up to stage 2b, before they were WCS aligned and combined with the \texttt{grizli}\footnote{\url{https://github.com/gbrammer/grizli/}} pipeline (see Brammer et al., in prep).
%
The images were pixel aligned at 40mas pixel scales, before producing a multi-wavelength catalog with \texttt{SExtractor} \citep{bertin96}. Fluxes are measured in small circular apertures of 0\farcs32 diameter and corrected to total fluxes using the AUTO fluxes from the F200W detection image. 

To determine stellar masses, SED fits to the 7 bands were done using the latest versions of the CIGALE and Prospector codes \citep{Boquien-M.2019CIGALE:-a-pytho,Johnson2021Stellar-Populat}, exploring a relatively wide range of priors.
Prior to correction for magnification, we find masses of $\log(\mstar/\msun) \sim 8.9-9.2$ and significant  
differences between the codes. We also note that the masses derived by \cite{Carnall2022A-first-look-at} are systematically lower than ours, which should mainly be due to consistently younger ages found by these authors.
Before more detailed analyses of the SEDs and spectra of these galaxies become available, we  adopt the stellar masses from our preferred CIGALE fits and a conservative uncertainty of $\pm 0.3$ (see Table 1).
To correct the  masses for gravitational magnification from the cluster, we use the magnification factors $\mu$ from the lens models of \cite{Caminha2022First-JWST-obse}. 
Other lens models, e.g.\ those used by \cite{Carnall2022A-first-look-at}, yield similar magnifications
(differences of 10-50\% at most) for the sources studied here.

\subsection{Comparison samples}

For comparison with low-$z$ galaxies we use a sample of 5607 star-forming galaxies from the SDSS Data Release 14
compiled by Y.\ Izotov and collaborators, analysed in earlier publications \citep[e.g.][]{Guseva2019Mg-II-lambda279,Ramambason2020Reconciling-esc}. 
The selection criteria used for the extraction of star-forming galaxies are presented in \cite{Izotov2014Multi-wavelengt}. 
Then we require a detection of the \Oiiit\ line with an accuracy better than $4 \sigma$, allowing thus direct 
abundance determinations using the $T_e$-method. Subsequently we refer to this sample as Izotov-DR14.

We also use the spectra of 89 galaxies at $z \sim 0.3$ from the Low-Z Lyman Continuum Survey (LzLCS), the first large sample
of galaxies with UV spectroscopy covering both the Lyman continuum and non-ionizing UV \citep[see][]{Flury2022aThe-Low-redshif,Flury2022bThe-Low-redshif}.
Approximately half of the sample has \Oiiit\ detections.

\begin{figure*}[htb]
{\centering
\includegraphics[width=8.5cm]{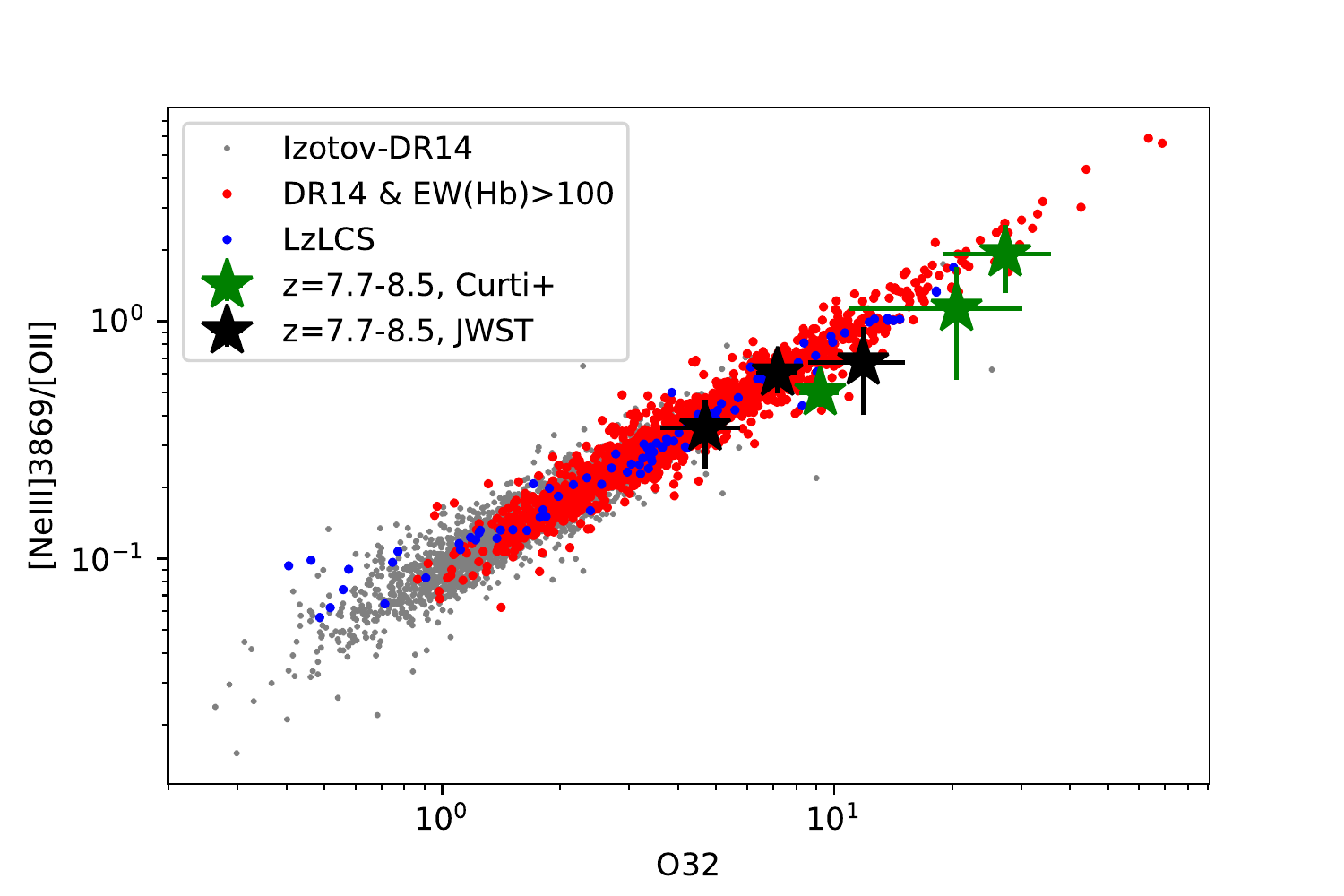}
\includegraphics[width=8.5cm]{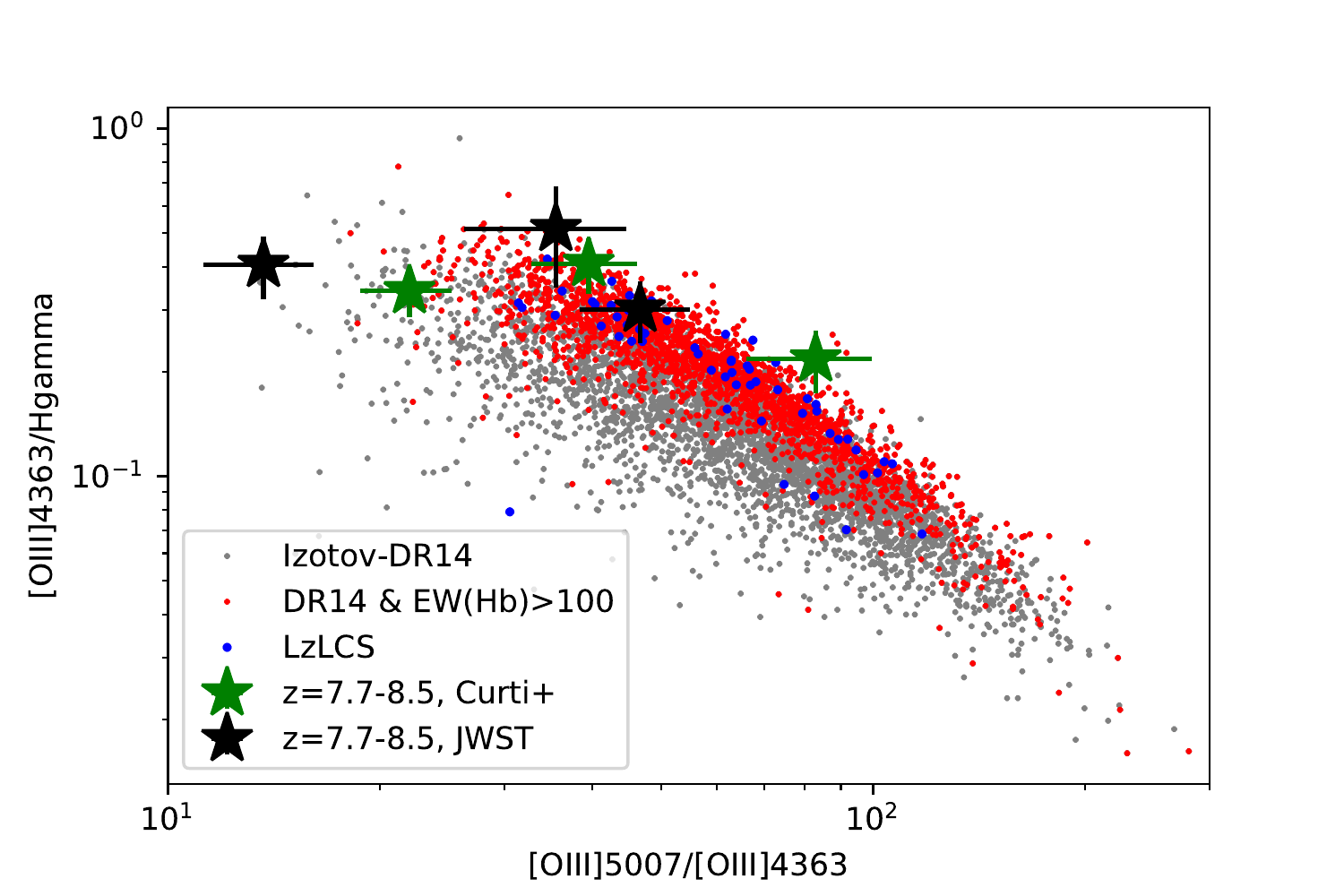}
\caption{Observed Ne3O2 ratio versus O32 (left panel) and \Oiiit/\hg\ versus \Oiiib/\Oiiit\ (right panel) for the $z \sim 8$ galaxies (black stars: our measurements; green stars: data from \protect\cite{Curti2022The-chemical-en}), 
the Izotov-DR14 sample of star-forming galaxies (grey and red points, according to EW(\hb)$<100$ \AA\ and $>100$ \AA, respectively), and the LzLCS sample (blue points). }}
\label{fig_1}
\end{figure*}

\begin{figure*}[htb]
{\centering
\includegraphics[width=8.5cm]{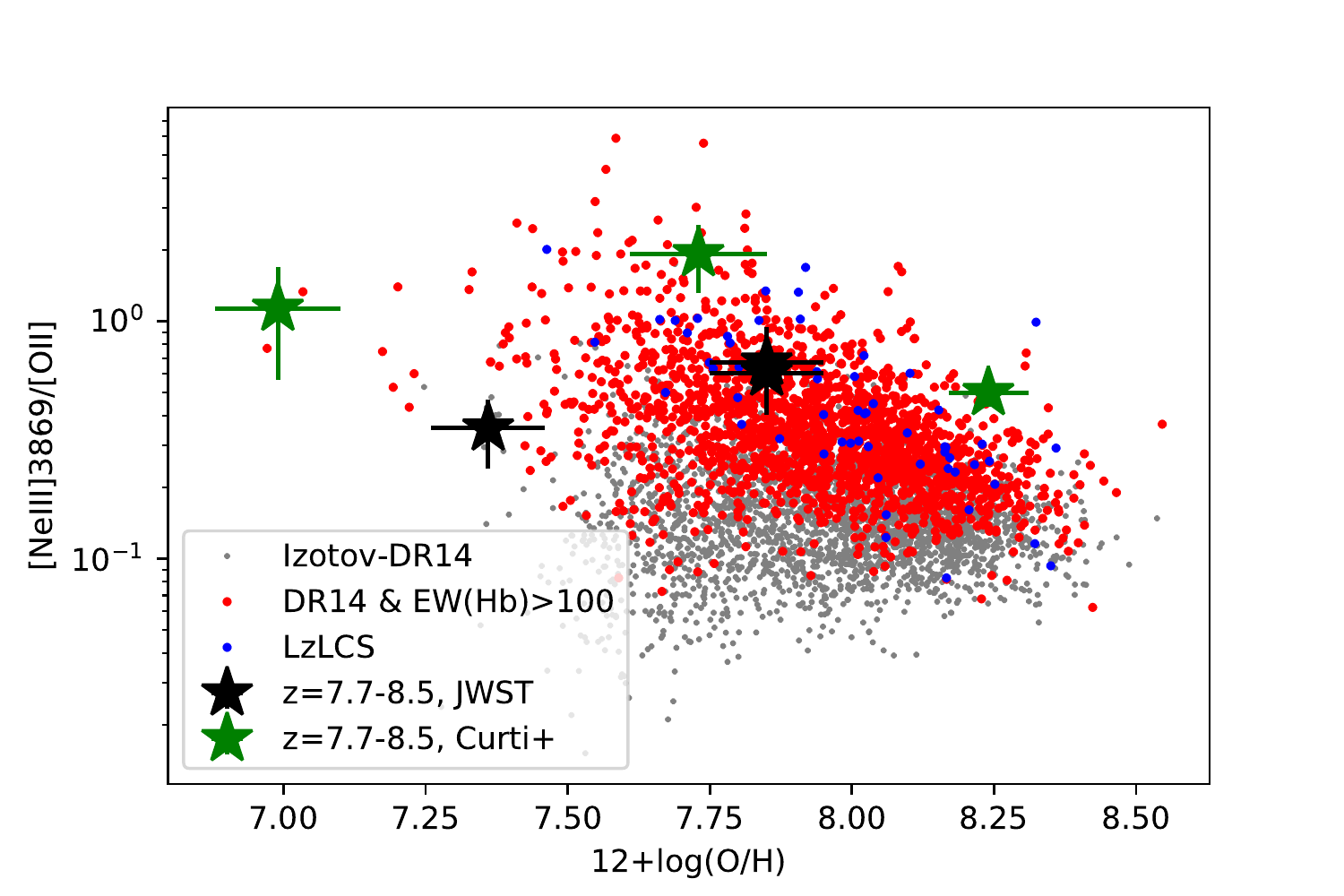}
\includegraphics[width=8.5cm]{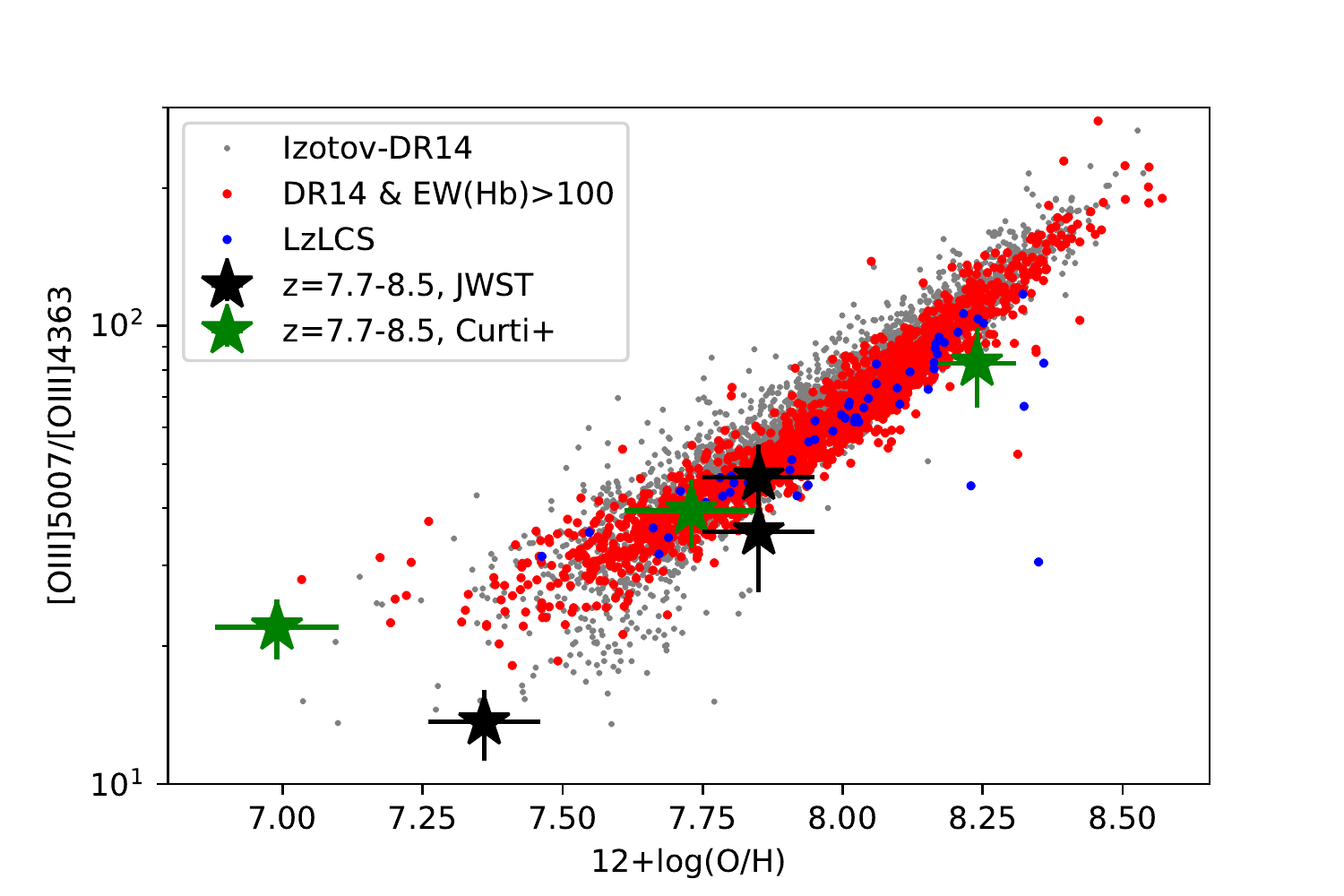}
\caption{\Neiii/\Oii\ ratio (left panel) and intensity of the auroral \Oiiit\ line relative to \Oiiib\ (right panel), as a function of O/H.  Same symbols as in Fig.\ 3.}
}
\label{fig_2}
\end{figure*}

\begin{figure}[htb]
{\centering
\includegraphics[width=8.5cm]{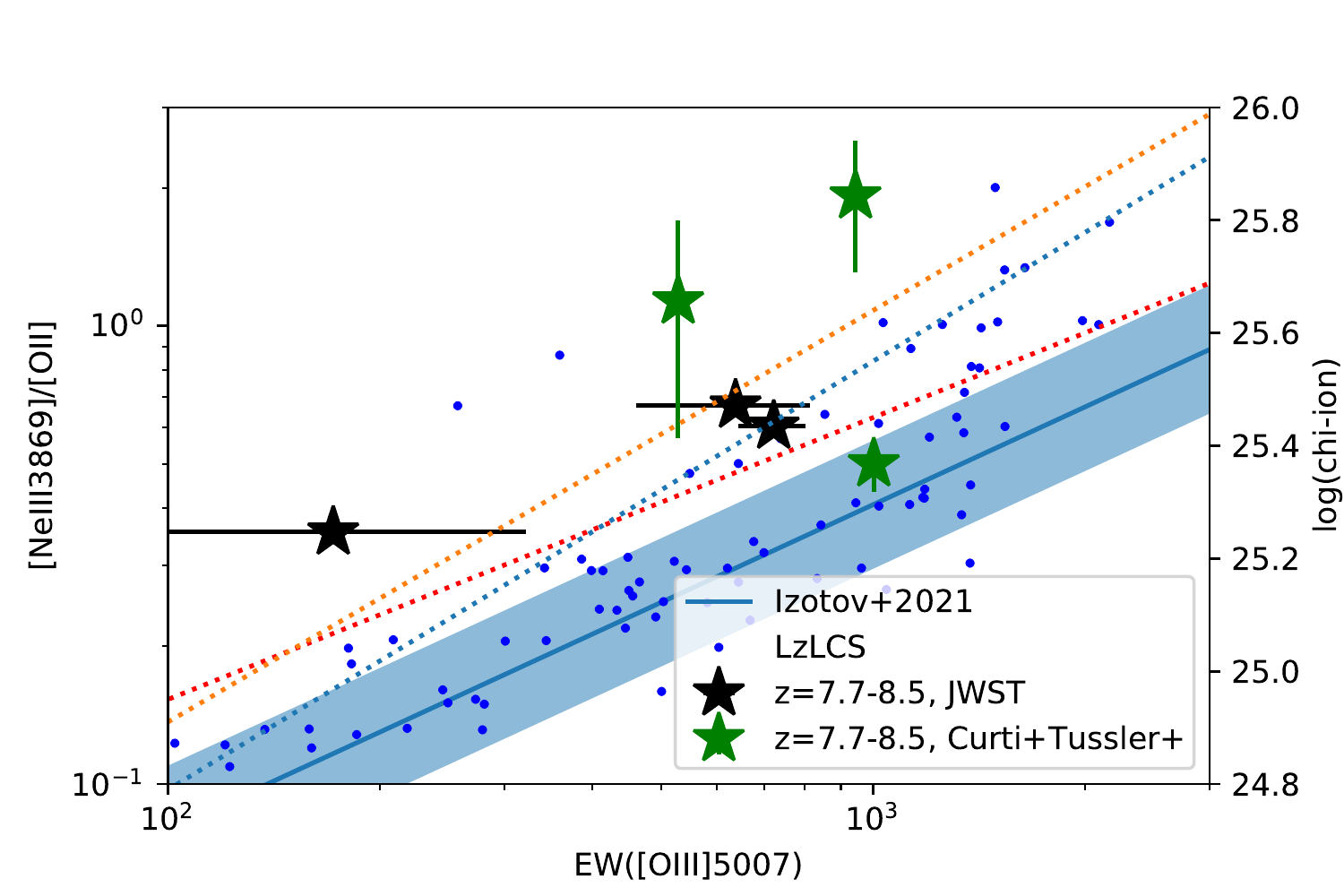}
\caption{Relation between the observed Ne3O2 ratio (left y-axis), \chion\ (right y-axis), and EW(\Oiiib). 
The $z \sim 8$ galaxies are marked with by black stars for our measurements and green stars using the line ratios and equivalent widths taken from 
\cite{Curti2022The-chemical-en} and \cite{Trussler2022Seeing-sharper-} respectively.
The shaded stripe shows the mean Ne3O2--EW relation and scatter in the large low-$z$ sample of \cite{Izotov2021Low-redshift-co}. 
The dotted lines show the relations between \chion\ (right axis) and EW(\Oiiib) found at $z \sim 0-1$ \citep{Chevallard2018Physical-proper,Izotov2021Low-redshift-co}.
}
}
\label{fig_3}
\end{figure}

\begin{figure}[htb]
{\centering
\includegraphics[width=8.5cm]{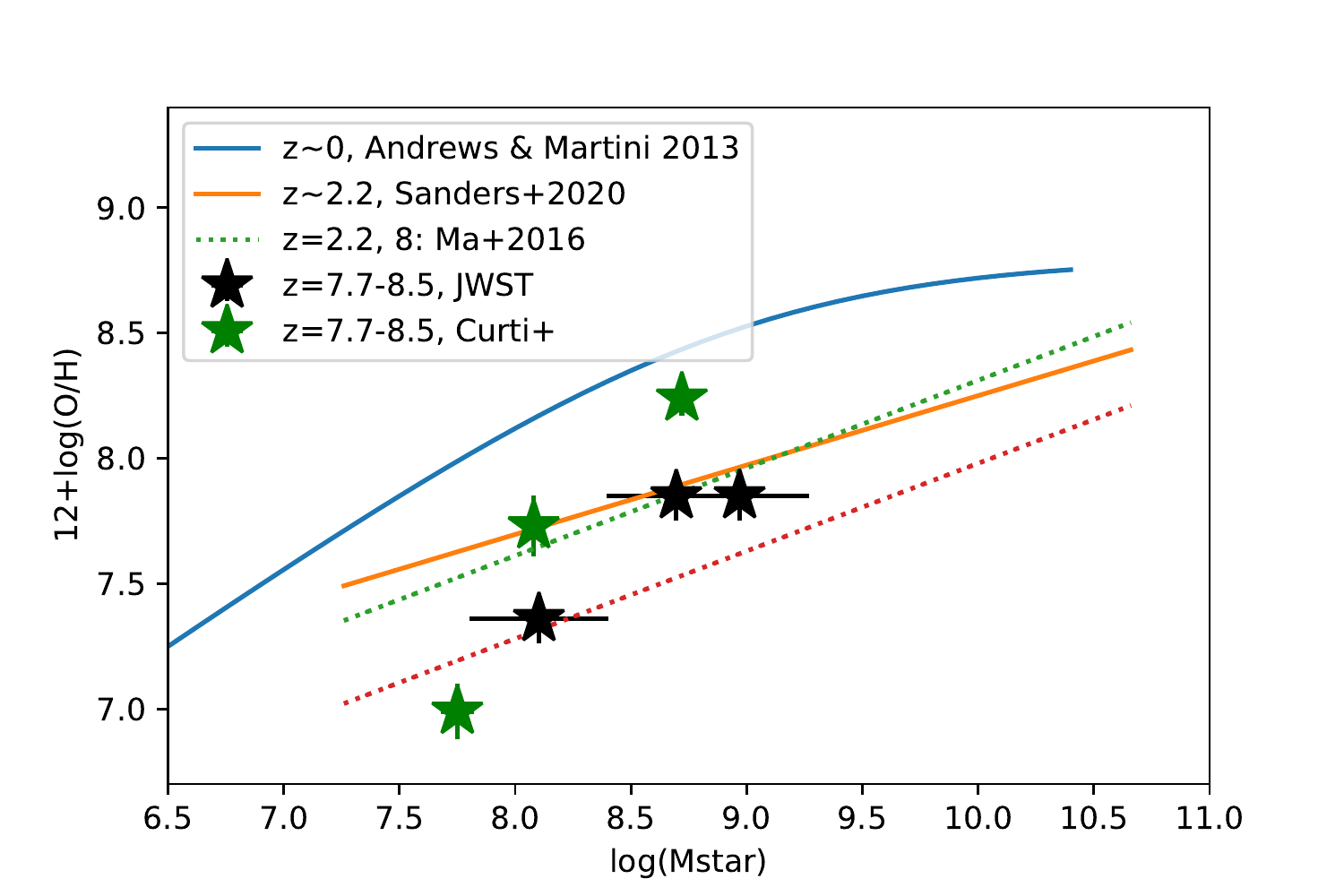}
\caption{Mass-metallicity relation of the three $z \sim 8$ galaxies (black stars) compared to mean relations observed at
$z=0$ and 2.2 \citep{Andrews2013The-Mass-Metall,Sanders2020The-MOSDEF-surv}, and results from numerical simulations at $z=2.2$ and 8 (dotted lines) from \cite{Ma2016The-origin-and-}.
Green stars show the results from  adopting stellar masses determined by \protect\cite{Curti2022The-chemical-en}, who quote very small uncertaintes for their stellar masses.}
}
\label{fig_4}
\end{figure}


\section{Observed and derived properties of the $z \sim 8$ galaxies}
\label{s_discuss}

 \subsection{Emission line properties} 
First we examine the observed emission line ratios in the three high-$z$ galaxies and compare them to those of
the low-$z$ samples (cf.\ above).
Figure 3 
shows the main line ratios including the lines of \Neiii, \Oii, \Oiiit, \Oiiib, and H$\gamma$, which are
detected in the JWST spectra.
The \neiii/\oii\ (Ne3O2) ratio is well known to closely trace O32,
since both high ionization lines of \neiii\ and \oiii\ originate in the same zone of the \hii\ region. 
Our measurements in the high-$z$ galaxies \citep[and those from][shown for comparison]{Curti2022The-chemical-en}  are compatible with the observed correlation, providing  confidence for our empirical flux calibration. 
By these line ratios, the three high-$z$ galaxies are found at relatively high excitation,
corresponding to the tail of the distribution in low-$z$ compact star-forming galaxies \citep[cf.][]{Izotov2021Low-redshift-co}.
Compared to the low-$z$ samples, the intensity of \Oiiit\ is also found to be relatively high, with \Oiiit$/ \hg \sim 0.3-0.5$, 
as illustrated in Fig.\ 3. 
The highest-$z$ source (04590) is a bit offset towards low \Oiiib/\Oiiit. 

The detection of the auroral \Oiiit\ line and \Oiiib\ provides access to the electron temperature $T_e$, and thus allows 
abundance determinations using the so-called ``direct method'' \citep[see e.g.][]{Kewley2002Using-Strong-Li}.
To do this we follow the prescriptions of \cite{Izotov2006The-chemical-co} assuming low densities. The results are listed in Table 1.
We find electron temperatures $T_e({\rm O}^{2+}) \sim 16000-18000$ K for the two $z=7.7$ galaxies, and nebular O/H abundances
of $\oh \approx 7.86-7.88$ for the two objects. This includes both ionic abundances of O$^+$ and O$^{2+}$, determined from the 
optical lines; in both cases  O$^{2+}$ dominates.
For 04590 the unusually low  \Oiiib/\Oiiit\ ratio leads to unphysically high electron temperatures ($T_e  \sim 35$ kK) and in excess of the 
highest accurate electron temperatures measured for star-forming galaxies at low-$z$ \citep[$T_e=24800 \pm 900$ K, ][]{Izotov2021J22292725:-an-e}.
This result is consistently found by all other papers who have analysed the JWST, and some authors have derived electron temperatures
as high as $T_e=37200 \pm 9900$ K \citep{Rhoads2022Finding-Peas-in}.
Although the detection of the \Oiiit\ line is beyond doubt, we consider that better data will be needed before the electron temperatures can
accurately be determined. Therefore the direct metallicity determinations for these high-$z$ galaxies should presently be taken with caution.

Different metallicity estimates can be obtained for the $z=8.495$ galaxy 04590 (see Table 1): 
From their finding of $T_e \ga 20$ kK, \cite{Trump2022The-Physical-Co} conclude that $\oh < 7.69$  ($<1/10$ solar), from the empirical relation
between $T_e$ and \oh\ \citep[cf.][]{Perez-Montero2021Extreme-emissio}.
Since standard strong line methods are not applicable at such low metallicities and for objects with ``extreme'' line ratios,
we can, e.g., use the empirical calibration of \cite{Izotov2021J22292725:-an-e}  based on measurements of R23\footnote{$R23=($\Oiii$/$\hb$+$\Oii$/$\hb$)$} and O32, and which has been established at $\oh <7.5$.
For 04590 we find $\oh = 7.39$ with our measurements, and $\oh = 7.24$ using the line ratios reported by \cite{Curti2022The-chemical-en}, who derived $\oh = 6.99 \pm 0.11$ by the direct method. 
With the measurements from \cite{Rhoads2022Finding-Peas-in} we obtain an intermediate metallicity, $\oh = 7.30$, using the same calibration.
All studies so far agree that  04590 shows the lowest metallicity among the three $z \sim 8$ galaxies.

In Fig.\ 4 
we show the Ne3O2 and  \Oiiit$/ \hg$ ratios as a function of metallicity for the low-$z$ samples and
the three $z \sim 8$ galaxies. We use these line ratios which are close in wavelengths to minimize possible uncertainties
of the flux calibration, differential slit losses and others.
Empirically, these line ratios can also provide a simple estimate of the metallicity O/H, as discussed by earlier studies
\citep[see e.g.][]{Nagao2006Gas-metallicity,Sanders2020The-MOSDEF-surv}.
In any case, we see that our metallicity estimates  of the high-$z$ galaxies lead to fairly compatible locations in these diagrams.
We conclude that the three $z \sim 8$ star-forming galaxies have low metallicities, in the range of $\oh \sim 7.4-8.0$.
More accurate determinations and a proper evaluation of the uncertainties await overall better data, including higher S/N data, proper calibrations, a more sophisticated data reduction, and continuum and line flux extractions.

Since the continuum is also detected in the NIRSpec spectra of the two $z =7.7$ galaxies, and very weakly so also in the third object,
we have also measured the \oiii\ equivalent width (see Table 1). 
We compare our measurements with those from the low-$z$ galaxies in Fig.\ 5,
where strong correlations between EW(\Oiiib)
and properties such as O32, Ne3O2, and others have been found \citep{Tang2019MMT/MMIRS-spect,Izotov2021Low-redshift-co}.
Possibly, the high-$z$ sources are somewhat offset. In any case, the EWs are high in the galaxies with significant continuum detections,
with EW(\Oiiib)$\sim 700$ \AA\ in the brightest source. 

Empirically, the ionizing photon production efficiency, \chion, is found to increase with the \Oiiib\ equivalent width, as also shown in Fig.\ 5. 
Using the relations found at low-$z$ and $z \sim 1-2$ \citep[see][]{Tang2019MMT/MMIRS-spect,Izotov2021Low-redshift-co}, 
we estimate $\log(\chion) \approx 25.1-25.5$ erg$^{-1}$ Hz, up to a factor $\sim 2$ higher than the ``canocical'' value often assumed
in ionizing photon budget calculations \citep{Robertson2013New-Constraints}. This is comparable to other estimates of \chion\ at high redshift
\citep[e.g.][]{Stefanon2022High-Equivalent}.
 
\begin{table}
\caption{Observed and derived quantities for the three high-$z$ galaxies}
\resizebox{0.47\textwidth}{!}{%
\begin{tabular}{lrrrrrr}
Quantity  & 04590 & 06355 & 10612 \\
\hline \\
redshift $z$ & 8.495 & 7.664 & 7.660 \\ 
$T_e$(OIII) & & 15584 & 18146 \\
\oh$^a$ &          & 7.86 & 7.88 \\
\oh$^b$  & 7.50 & 7.85 & 8.0   \\
\oh$^c$  & 7.39 & 7.76 & 7.91   \\ 
O32 & 7.0 & 6.3 & 10.6 \\
EW(\Oiiib) & $172 \pm 150$ & $723 \pm 78$ & $638 \pm 176$\\
$m_{F277W}$ & 27.8 & 26.6 & 25.6 \\
\muv$^d$ & $-20.29$ & $-21.09$ & $-20.38$ \\
\muv$^e$& $-18.06$ & $-20.51$ & $-19.81$ \\
$\beta_{1500}$ & $-2.20 \pm 0.15$ & $-1.96 \pm 0.22$ &  $-2.31 \pm 0.11$ \\
$\log(\mstar) \times \mu$ & $9.0 \pm 0.3$ & $9.2 \pm 0.3$ & $8.9 \pm 0.3$ \\
Magnification $\mu$ & 7.9 & 1.7 & 1.7 \\ 
\hline \\
\multicolumn{4}{l}{$^a$ direct method, $^b$ assuming $T_e=16000$ K}\\
\multicolumn{4}{l}{$^c$ using strong line methods (Izotov+2019,2021)} \\ 
\multicolumn{4}{l}{$^d$ observed, including lensing}\\
\multicolumn{4}{l}{$^e$ corrected for lensing}\\\end{tabular}}
\label{tab_1}
\end{table}

\subsection{Physical properties of the $z \sim 8$ galaxies and the mass-metallicity relation}

As discussed, the three $z \sim 8$ galaxies show emission line properties comparable to 
compact star-forming galaxies at low-$z$ with strong emission lines. 
\cite{Izotov2021Low-redshift-co} have shown that the low-$z$ galaxies with strong lines (EW(\hb)$>100$ \AA) 
are good analogues of many of the $z \sim 1-3$ star-forming galaxies (Lyman alpha emitters and Lyman break galaxies)
studied so far. By inference the emission line properties of the three $z \sim 8$ galaxies studied here 
therefore also resemble those at intermediate redshifts.

By construction, the galaxies selected here cannot be claimed to be ``typical'', and larger, systematic studies
will be needed. 
From our measurements, one object (10612) shows a very high ratio \Oiiib$/\hb \approx 10$
and a strong \Oiiit\ line, which could indicate an active galaxy \cite[Seyfert 2, see also][]{Brinchmann2022High-z-galaxies,Curti2022The-chemical-en}.
On the other hand, we clearly find evidence for one star-forming galaxy with a fairly low metallicity 
(04590 with $\oh \approx 7.2-7.4$).

If we combine our nebular metallicity estimates with the stellar masses described earlier, we obtain the 
mass-metallicity relation shown in Fig.\ 6. 
Our objects are found close to or below the mass-metallicity relation observed at $z \sim 2$ and possibly
offset by $\sim 0.2-0.3$ below this relation, in good agreement with the relation derived by \cite{Ma2016The-origin-and-}
from simulations. Similar results were also obtained by \cite{Jones2020The-Mass-Metall} using an indirect method
based on ALMA emission line detections.
However, at the present stage we consider the stellar masses uncertain, since these depend
significantly on assumptions of the star formation history and on the age of stellar populations \citep[see also][]{Tang2022Stellar-populat}, 
plus uncertainties with the zeropoints of NIRCam \citep[see][]{Rigby2022Characterizatio}.
For example, comparing the masses derived from SEDs by  \cite{Carnall2022A-first-look-at},
 \cite{Curti2022The-chemical-en}, and our work, we can see significant differences which are generally larger than
 the uncertainties cited.
Adopting the lower masses from \cite{Carnall2022A-first-look-at}, e.g.,  which would imply that our $z \sim 8$ galaxies lie 
close to the $z \sim 2$ mass-metallicity relation.
The lower masses are mostly due to the very young ages ($\sim 1-2$ Myr) inferred by these authors, whereas our SED fits favour less extreme populations.
Although \cite{Carnall2022A-first-look-at} claim that the SEDs show evidence for Balmer jumps (i.e.\ Balmer breaks
in emission due to strong nebular continuum), we do not see such a behaviour in the NIRSpec spectrum of the brightest
source, 06355, shown in Fig.\ \ref{fig_spec}.

Having shown that the $z \sim 8$ galaxies closely resemble strong emission line galaxies from our low-$z$ sample
it is tempting to infer indirectly other properties using correlations established at low-$z$.
Certainly, the high O32 ratios, low metallicity, and blue UV slopes ($\beta$) suggest that these galaxies could contribute
to cosmic reionization, i.e.\ have escaping ionizing photons. For example, for the observed values of O32, \oh, and $\beta$,
the LzLCS results suggest a 30-60\% detection fraction of the Lyman continuum. Adopting the mean relation between the 
LyC escape fraction, \fesc, and the UV slope, we estimate $\fesc = 0.03-0.08$
for the three $z \sim 8 $ galaxies, although the LyC escape could also be significantly higher \citep[see][]{Chisholm2022The-Far-Ultravi}.

Future improvements in the calibration and data reduction and additional observations will allow us to determine 
more accurately equivalent widths and total line fluxes of emission lines, measure continuum shapes, combine
photometry and spectra etc. and hence improve our knowledge of the physical properties of galaxies at high-redshift.

\section{Conclusion}
\label{s_conclude}
We have analysed the rest-frame optical spectra of two galaxies at $z=7.7$ and one at $z=8.5$ from the JWST Early Release Observations. 
The spectra, exhibit numerous emission lines of H, \hei, \oii, \oiii, and \neiii, as commonly seen in metal-poor star-forming galaxies at low redshift.
They provide, for the first time in the epoch of reionization, detailed information on the chemical composition and interstellar medium of these galaxies.
Our main results are summarized as follows:
\begin{itemize}
\item The auroral \Oiiit\ line is significantly detected in all galaxies with $3.9-5.7 \sigma$, allowing the determination of the O/H abundance (metallicity) using the direct method. 
With this method and different strong-line methods we estimate metallicities between $\oh=7.4-8.0$, i.e.\ $\sim 5-20$ \% of solar.
\item  All three galaxies show a high excitation, as measured by their line ratios of O32$=6-11$ and Ne3O2$=0.4-0.7$, or even higher, according 
to \cite{Curti2022The-chemical-en} and \cite{Rhoads2022Finding-Peas-in}.
The observed emission line ratios are similar to those of rare low-$z$ star-forming galaxies, which are considered analogues of high-redshift
($z \sim 1-3$) galaxies. One of the $z=7.6$ galaxies shows unusually high \Oiiib/\hb$ \approx 10$, possibly indicative of nuclear activity (Seyfert 2).
\item The $z \sim 8$ galaxies show quite high equivalent widths, e.g.\ EW(\Oiiib) up to $\sim 700$ \AA, as expected from low-$z$ galaxies with high excitation.
Such galaxies are known to be efficient producers of ionizing photons. We conservatively estimate $\log(\chion) \sim 25.2-25.5$ erg$^{-1}$ Hz for our galaxies.
\item Using stellar mass estimates from SED fits, we find the $z \sim 8$ galaxies to lie close to or below the mass-metallicity relation (MZR) at $z \sim 2$.
To assess if the MZR continues to evolve from $z=2$ to 8 will need future studies of larger galaxy samples and accurate metallicity determinations.
\end{itemize}

Overall, the first analysis of the rest-frame optical spectra of galaxies at $z \sim 8$ indicates that the emission lines properties
of galaxies in the epoch of reionization resemble those of relatively rare ``local analogues'' previously studied from the SDSS,
and for which numerous physical properties have already been determined. 
These low-$z$ samples will soon be rivaled by numerous measurements with NIRSpec  onboard JWST.
Clearly, the first data release reveals already an extremely promising ``preview'' of upcoming science with JWST in the early Universe. 
More robust inferences will require better data, including higher S/N spectra and improvements in the calibration, data reduction, and continuum and line flux extractions.

\begin{acknowledgements}
We are very grateful to the ERO and other teams who made these JWST observations possible.
The Early Release Observations and associated materials were developed, executed, and compiled by the ERO production team:  Hannah Braun, Claire Blome, Matthew Brown, Margaret Carruthers, Dan Coe, Joseph DePasquale, Nestor Espinoza, Macarena Garcia Marin, Karl Gordon, Alaina Henry, Leah Hustak, Andi James, Ann Jenkins, Anton Koekemoer, Stephanie LaMassa, David Law, Alexandra Lockwood, Amaya Moro-Martin, Susan Mullally, Alyssa Pagan, Dani Player, Klaus Pontoppidan, Charles Proffitt, Christine Pulliam, Leah Ramsay, Swara Ravindranath, Neill Reid, Massimo Robberto, Elena Sabbi, Leonardo Ubeda. The EROs were also made possible by the foundational efforts and support from the JWST instruments, STScI planning and scheduling, and Data Management teams.
We acknowledge support from the Swiss National Science Foundation through project grant 200020\_207349 (PAO, LB).
The Cosmic Dawn Center (DAWN) is funded by the Danish National Research Foundation under grant No.\ 140.
Cloud-based data processing and file storage for this work is provided by the AWS Cloud Credits for Research program. 
Y.I. and N.G. acknowledge support from the National Academy of Sciences of Ukraine 
by its priority project No.\ 0122U002259 “Fundamental properties of the matter and its manifestation 
in micro world, astrophysics and cosmology”.

\end{acknowledgements}
\bibliographystyle{aa}
\bibliography{merge_misc_highz_literature}

\begin{thebibliography}{40}
\expandafter\ifx\csname natexlab\endcsname\relax\def\natexlab#1{#1}\fi

\bibitem[{{Ahumada} {et~al.}(2020){Ahumada}, {Prieto}, {Almeida}, {Anders},
  {Anderson}, {Andrews}, {Anguiano}, {Arcodia}, {Armengaud}, {Aubert}, {Avila},
  {Avila-Reese}, {Badenes}, {Balland}, {Barger}, {Barrera-Ballesteros}, {Basu},
  {Bautista}, {Beaton}, {Beers}, {Benavides}, {Bender}, {Bernardi}, {Bershady},
  {Beutler}, {Bidin}, {Bird}, {Bizyaev}, {Blanc}, {Blanton}, {Boquien},
  {Borissova}, {Bovy}, {Brandt}, {Brinkmann}, {Brownstein}, {Bundy}, {Bureau},
  {Burgasser}, {Burtin}, {Cano-D{\'\i}az}, {Capasso}, {Cappellari}, {Carrera},
  {Chabanier}, {Chaplin}, {Chapman}, {Cherinka}, {Chiappini}, {Doohyun Choi},
  {Chojnowski}, {Chung}, {Clerc}, {Coffey}, {Comerford}, {Comparat}, {da
  Costa}, {Cousinou}, {Covey}, {Crane}, {Cunha}, {Ilha}, {Dai}, {Damsted},
  {Darling}, {Davidson}, {Davies}, {Dawson}, {De}, {de la Macorra}, {De Lee},
  {Queiroz}, {Deconto Machado}, {de la Torre}, {Dell'Agli}, {du Mas des
  Bourboux}, {Diamond-Stanic}, {Dillon}, {Donor}, {Drory}, {Duckworth},
  {Dwelly}, {Ebelke}, {Eftekharzadeh}, {Davis Eigenbrot}, {Elsworth},
  {Eracleous}, {Erfanianfar}, {Escoffier}, {Fan}, {Farr},
  {Fern{\'a}ndez-Trincado}, {Feuillet}, {Finoguenov}, {Fofie},
  {Fraser-McKelvie}, {Frinchaboy}, {Fromenteau}, {Fu}, {Galbany}, {Garcia},
  {Garc{\'\i}a-Hern{\'a}ndez}, {Oehmichen}, {Ge}, {Maia}, {Geisler}, {Gelfand},
  {Goddy}, {Gonzalez-Perez}, {Grabowski}, {Green}, {Grier}, {Guo}, {Guy},
  {Harding}, {Hasselquist}, {Hawken}, {Hayes}, {Hearty}, {Hekker}, {Hogg},
  {Holtzman}, {Horta}, {Hou}, {Hsieh}, {Huber}, {Hunt}, {Chitham}, {Imig},
  {Jaber}, {Angel}, {Johnson}, {Jones}, {J{\"o}nsson}, {Jullo}, {Kim},
  {Kinemuchi}, {Kirkpatrick}, {Kite}, {Klaene}, {Kneib}, {Kollmeier}, {Kong},
  {Kounkel}, {Krishnarao}, {Lacerna}, {Lan}, {Lane}, {Law}, {Le Goff}, {Leung},
  {Lewis}, {Li}, {Lian}, {Lin}, {Long}, {Longa-Pe{\~n}a}, {Lundgren}, {Lyke},
  {Ted Mackereth}, {MacLeod}, {Majewski}, {Manchado}, {Maraston}, {Martini},
  {Masseron}, {Masters}, {Mathur}, {McDermid}, {Merloni}, {Merrifield},
  {M{\'e}sz{\'a}ros}, {Miglio}, {Minniti}, {Minsley}, {Miyaji}, {Mohammad},
  {Mosser}, {Mueller}, {Muna}, {Mu{\~n}oz-Guti{\'e}rrez}, {Myers}, {Nadathur},
  {Nair}, {Nandra}, {do Nascimento}, {Nevin}, {Newman}, {Nidever}, {Nitschelm},
  {Noterdaeme}, {O'Connell}, {Olmstead}, {Oravetz}, {Oravetz}, {Osorio},
  {Pace}, {Padilla}, {Palanque-Delabrouille}, {Palicio}, {Pan}, {Pan},
  {Parker}, {Paviot}, {Peirani}, {Ram{\'r}ez}, {Penny}, {Percival},
  {Perez-Fournon}, {P{\'e}rez-R{\`a}fols}, {Petitjean}, {Pieri},
  {Pinsonneault}, {Poovelil}, {Povick}, {Prakash}, {Price-Whelan}, {Raddick},
  {Raichoor}, {Ray}, {Rembold}, {Rezaie}, {Riffel}, {Riffel}, {Rix}, {Robin},
  {Roman-Lopes}, {Rom{\'a}n-Z{\'u}{\~n}iga}, {Rose}, {Ross}, {Rossi},
  {Rowlands}, {Rubin}, {Salvato}, {S{\'a}nchez}, {S{\'a}nchez-Menguiano},
  {S{\'a}nchez-Gallego}, {Sayres}, {Schaefer}, {Schiavon}, {Schimoia},
  {Schlafly}, {Schlegel}, {Schneider}, {Schultheis}, {Schwope}, {Seo},
  {Serenelli}, {Shafieloo}, {Shamsi}, {Shao}, {Shen}, {Shetrone}, {Shirley},
  {Aguirre}, {Simon}, {Skrutskie}, {Slosar}, {Smethurst}, {Sobeck}, {Sodi},
  {Souto}, {Stark}, {Stassun}, {Steinmetz}, {Stello}, {Stermer},
  {Storchi-Bergmann}, {Streblyanska}, {Stringfellow}, {Stutz}, {Su{\'a}rez},
  {Sun}, {Taghizadeh-Popp}, {Talbot}, {Tayar}, {Thakar}, {Theriault}, {Thomas},
  {Thomas}, {Tinker}, {Tojeiro}, {Toledo}, {Tremonti}, {Troup}, {Tuttle},
  {Unda-Sanzana}, {Valentini}, {Vargas-Gonz{\'a}lez}, {Vargas-Maga{\~n}a},
  {V{\'a}zquez-Mata}, {Vivek}, {Wake}, {Wang}, {Weaver}, {Weijmans}, {Wild},
  {Wilson}, {Wilson}, {Wolthuis}, {Wood-Vasey}, {Yan}, {Yang}, {Y{\`e}che},
  {Zamora}, {Zarrouk}, {Zasowski}, {Zhang}, {Zhao}, {Zhao}, {Zheng}, {Zheng},
  {Zhu}, \& {Zou}}]{Ahumada2020The-16th-Data-R}
{Ahumada}, R., {Prieto}, C.~A., {Almeida}, A., {et~al.} 2020, \apjs, 249, 3

\bibitem[{{Andrews} \& {Martini}(2013)}]{Andrews2013The-Mass-Metall}
{Andrews}, B.~H. \& {Martini}, P. 2013, \apj, 765, 140

\bibitem[{{Arellano-C{\'o}rdova} {et~al.}(2022){Arellano-C{\'o}rdova}, {Berg},
  {Chisholm}, {Arrabal Haro}, {Dickinson}, {Finkelstein}, {Leclercq}, {Rogers},
  {Simons}, {Skillman}, \& {Trump}}]{Arellano-Cordova2022A-First-Look-at}
{Arellano-C{\'o}rdova}, K.~Z., {Berg}, D.~A., {Chisholm}, J., {et~al.} 2022,
  arXiv e-prints, arXiv:2208.02562

\bibitem[{{Bertin} \& {Arnouts}(1996)}]{bertin96}
{Bertin}, E. \& {Arnouts}, S. 1996, \aaps, 117, 393

\bibitem[{{Boquien} {et~al.}(2019){Boquien}, {Burgarella, D.}, {Roehlly, Y.},
  {Buat, V.}, {Ciesla, L.}, {Corre, D.}, {Inoue, A. K.}, \& {Salas,
  H.}}]{Boquien-M.2019CIGALE:-a-pytho}
{Boquien}, M., {Burgarella, D.}, {Roehlly, Y.}, {et~al.} 2019, A\&A, 622, A103

\bibitem[{{Brammer}(2022)}]{Brammer2022Images-and-cata}
{Brammer}, G. 2022, {Images and catalogs of HST and JWST images in the
  SMACS-0723 field}

\bibitem[{{Brinchmann}(2022)}]{Brinchmann2022High-z-galaxies}
{Brinchmann}, J. 2022, arXiv e-prints, arXiv:2208.07467

\bibitem[{{Caminha} {et~al.}(2022){Caminha}, {Suyu}, {Mercurio}, {Brammer},
  {Bergamini}, {Vanzella}, \& {Acebron}}]{Caminha2022First-JWST-obse}
{Caminha}, G.~B., {Suyu}, S.~H., {Mercurio}, A., {et~al.} 2022, arXiv e-prints,
  arXiv:2207.07567

\bibitem[{{Carnall} {et~al.}(2022){Carnall}, {Begley}, {McLeod}, {Hamadouche},
  {Donnan}, {McLure}, {Dunlop}, {Bondestam}, {Cullen}, {Jewell}, \&
  {Pollock}}]{Carnall2022A-first-look-at}
{Carnall}, A.~C., {Begley}, R., {McLeod}, D.~J., {et~al.} 2022, arXiv e-prints,
  arXiv:2207.08778

\bibitem[{{Chevallard} {et~al.}(2018){Chevallard}, {Charlot}, {Senchyna},
  {Stark}, {Vidal-Garc{\'{\i}}a}, {Feltre}, {Gutkin}, {Jones}, {Mainali}, \&
  {Wofford}}]{Chevallard2018Physical-proper}
{Chevallard}, J., {Charlot}, S., {Senchyna}, P., {et~al.} 2018, \mnras, 479,
  3264

\bibitem[{{Chisholm} {et~al.}(2022){Chisholm}, {Saldana-Lopez}, {Flury},
  {Schaerer}, {Jaskot}, {Amorin}, {Atek}, {Finkelstein}, {Fleming}, {Ferguson},
  {Fernandez}, {Giavalisco}, {Hayes}, {Heckman}, {Henry}, {Ji},
  {Marques-Chaves}, {Mauerhofer}, {McCandliss}, {Oey}, {Ostlin}, {Rutkowski},
  {Scarlata}, {Thuan}, {Trebitsch}, {Wang}, {Worseck}, \&
  {Xu}}]{Chisholm2022The-Far-Ultravi}
{Chisholm}, J., {Saldana-Lopez}, A., {Flury}, S., {et~al.} 2022, arXiv
  e-prints, arXiv:2207.05771

\bibitem[{{Curti} {et~al.}(2022){Curti}, {D'Eugenio}, {Carniani}, {Maiolino},
  {Sandles}, {Witstok}, {Baker}, {Bennett}, {Piotrowska}, {Tacchella},
  {Charlot}, {Nakajima}, {Maheson}, {Mannucci}, {Arribas}, {Belfiore},
  {Bonaventura}, {Bunker}, {Chevallard}, {Cresci}, {Curtis-Lake},
  {Hayden-Pawson}, {Kumari}, {Laseter}, {Looser}, {Marconi}, {Maseda}, {Jones},
  {Scholtz}, {Smit}, {Ubler}, \& {Wallace}}]{Curti2022The-chemical-en}
{Curti}, M., {D'Eugenio}, F., {Carniani}, S., {et~al.} 2022, arXiv e-prints,
  arXiv:2207.12375

\bibitem[{{Flury} {et~al.}(2022{\natexlab{a}}){Flury}, {Jaskot}, {Ferguson},
  {Worseck}, {Makan}, {Chisholm}, {Saldana-Lopez}, {Schaerer}, {McCandliss},
  {Wang}, {Ford}, {Heckman}, {Ji}, {Giavalisco}, {Amorin}, {Atek}, {Blaizot},
  {Borthakur}, {Carr}, {Castellano}, {Cristiani}, {De Barros}, {Dickinson},
  {Finkelstein}, {Fleming}, {Fontanot}, {Garel}, {Grazian}, {Hayes}, {Henry},
  {Mauerhofer}, {Micheva}, {Oey}, {Ostlin}, {Papovich}, {Pentericci},
  {Ravindranath}, {Rosdahl}, {Rutkowski}, {Santini}, {Scarlata}, {Teplitz},
  {Thuan}, {Trebitsch}, {Vanzella}, {Verhamme}, \&
  {Xu}}]{Flury2022aThe-Low-redshif}
{Flury}, S.~R., {Jaskot}, A.~E., {Ferguson}, H.~C., {et~al.}
  2022{\natexlab{a}}, \apjs, 260, 1

\bibitem[{{Flury} {et~al.}(2022{\natexlab{b}}){Flury}, {Jaskot}, {Ferguson},
  {Worseck}, {Makan}, {Chisholm}, {Saldana-Lopez}, {Schaerer}, {McCandliss},
  {Xu}, {Wang}, {Oey}, {Ford}, {Heckman}, {Ji}, {Giavalisco}, {Amor{\'\i}n},
  {Atek}, {Blaizot}, {Borthakur}, {Carr}, {Castellano}, {Barros}, {Dickinson},
  {Finkelstein}, {Fleming}, {Fontanot}, {Garel}, {Grazian}, {Hayes}, {Henry},
  {Mauerhofer}, {Micheva}, {Ostlin}, {Papovich}, {Pentericci}, {Ravindranath},
  {Rosdahl}, {Rutkowski}, {Santini}, {Scarlata}, {Teplitz}, {Thuan},
  {Trebitsch}, {Vanzella}, \& {Verhamme}}]{Flury2022bThe-Low-redshif}
{Flury}, S.~R., {Jaskot}, A.~E., {Ferguson}, H.~C., {et~al.}
  2022{\natexlab{b}}, \apj, 930, 126

\bibitem[{{F{\"o}rster Schreiber} \&
  {Wuyts}(2020)}]{Forster-Schreiber2020Star-Forming-Ga}
{F{\"o}rster Schreiber}, N.~M. \& {Wuyts}, S. 2020, \araa, 58, 661

\bibitem[{{Guseva} {et~al.}(2019){Guseva}, {Izotov}, {Fricke}, \&
  {Henkel}}]{Guseva2019Mg-II-lambda279}
{Guseva}, N.~G., {Izotov}, Y.~I., {Fricke}, K.~J., \& {Henkel}, C. 2019, \aap,
  624, A21

\bibitem[{{Izotov} {et~al.}(2014){Izotov}, {Guseva}, {Fricke}, \&
  {Henkel}}]{Izotov2014Multi-wavelengt}
{Izotov}, Y.~I., {Guseva}, N.~G., {Fricke}, K.~J., \& {Henkel}, C. 2014, \aap,
  561, A33

\bibitem[{{Izotov} {et~al.}(2021{\natexlab{a}}){Izotov}, {Guseva}, {Fricke},
  {Henkel}, {Schaerer}, \& {Thuan}}]{Izotov2021Low-redshift-co}
{Izotov}, Y.~I., {Guseva}, N.~G., {Fricke}, K.~J., {et~al.} 2021{\natexlab{a}},
  \aap, 646, A138

\bibitem[{{Izotov} {et~al.}(2006){Izotov}, {Stasi{\'n}ska}, {Meynet}, {Guseva},
  \& {Thuan}}]{Izotov2006The-chemical-co}
{Izotov}, Y.~I., {Stasi{\'n}ska}, G., {Meynet}, G., {Guseva}, N.~G., \&
  {Thuan}, T.~X. 2006, \aap, 448, 955

\bibitem[{{Izotov} {et~al.}(2021{\natexlab{b}}){Izotov}, {Thuan}, \&
  {Guseva}}]{Izotov2021J22292725:-an-e}
{Izotov}, Y.~I., {Thuan}, T.~X., \& {Guseva}, N.~G. 2021{\natexlab{b}}, \mnras,
  504, 3996

\bibitem[{{Johnson} {et~al.}(2021){Johnson}, {Leja}, {Conroy}, \&
  {Speagle}}]{Johnson2021Stellar-Populat}
{Johnson}, B.~D., {Leja}, J., {Conroy}, C., \& {Speagle}, J.~S. 2021, \apjs,
  254, 22

\bibitem[{{Jones} {et~al.}(2020){Jones}, {Sanders}, {Roberts-Borsani}, {Ellis},
  {Laporte}, {Treu}, \& {Harikane}}]{Jones2020The-Mass-Metall}
{Jones}, T., {Sanders}, R., {Roberts-Borsani}, G., {et~al.} 2020, \apj, 903,
  150

\bibitem[{{Katz} {et~al.}(2022){Katz}, {Saxena}, {Cameron}, {Carniani},
  {Bunker}, {Arribas}, {Bhatawdekar}, {Bowler}, {Boyett}, {Cresci},
  {Curtis-Lake}, {D'Eugenio}, {Kumari}, {Looser}, {Ubler}, {Willott}, \&
  {Witstok}}]{Katz2022First-Insights-}
{Katz}, H., {Saxena}, A., {Cameron}, A.~J., {et~al.} 2022, arXiv e-prints,
  arXiv:2207.13693

\bibitem[{{Kewley} \& {Dopita}(2002)}]{Kewley2002Using-Strong-Li}
{Kewley}, L.~J. \& {Dopita}, M.~A. 2002, \apjs, 142, 35

\bibitem[{Kewley {et~al.}(2019)Kewley, Nicholls, \&
  Sutherland}]{Kewley2019Understanding-G}
Kewley, L.~J., Nicholls, D.~C., \& Sutherland, R.~S. 2019, \araa, 57, 511

\bibitem[{Kriek {et~al.}(2015)Kriek, Shapley, Reddy, Siana, Coil, Mobasher,
  Freeman, de~Groot, Price, Sanders, Shivaei, Brammer, Momcheva, Skelton, van
  Dokkum, Whitaker, Aird, Azadi, Kassis, Bullock, Conroy, Dav{\'{e}},
  Kere{\v{s}}, \& Krumholz}]{Kriek2015THE-MOSFIRE-DEE}
Kriek, M., Shapley, A.~E., Reddy, N.~A., {et~al.} 2015, \apjs, 218, 15

\bibitem[{{Ma} {et~al.}(2016){Ma}, {Hopkins}, {Faucher-Gigu{\`e}re}, {Zolman},
  {Muratov}, {Kere{\v{s}}}, \& {Quataert}}]{Ma2016The-origin-and-}
{Ma}, X., {Hopkins}, P.~F., {Faucher-Gigu{\`e}re}, C.-A., {et~al.} 2016,
  \mnras, 456, 2140

\bibitem[{{Nagao} {et~al.}(2006){Nagao}, {Maiolino}, \&
  {Marconi}}]{Nagao2006Gas-metallicity}
{Nagao}, T., {Maiolino}, R., \& {Marconi}, A. 2006, \aap, 459, 85

\bibitem[{{P{\'e}rez-Montero} {et~al.}(2021){P{\'e}rez-Montero}, {Amor{\'\i}n},
  {S{\'a}nchez Almeida}, {V{\'\i}lchez}, {Garc{\'\i}a-Benito}, \&
  {Kehrig}}]{Perez-Montero2021Extreme-emissio}
{P{\'e}rez-Montero}, E., {Amor{\'\i}n}, R., {S{\'a}nchez Almeida}, J., {et~al.}
  2021, \mnras, 504, 1237

\bibitem[{{Ramambason} {et~al.}(2020){Ramambason}, {Schaerer}, {Stasi{\'n}ska},
  {Izotov}, {Guseva}, {V{\'\i}lchez}, {Amor{\'\i}n}, \&
  {Morisset}}]{Ramambason2020Reconciling-esc}
{Ramambason}, L., {Schaerer}, D., {Stasi{\'n}ska}, G., {et~al.} 2020, \aap,
  644, A21

\bibitem[{{Rhoads} {et~al.}(2022){Rhoads}, {Wold}, {Harish}, {Kim}, {Pharo},
  {Malhotra}, {Gabrielpillai}, {Jiang}, \& {Yang}}]{Rhoads2022Finding-Peas-in}
{Rhoads}, J.~E., {Wold}, I. G.~B., {Harish}, S., {et~al.} 2022, arXiv e-prints,
  arXiv:2207.13020

\bibitem[{{Rigby} {et~al.}(2022){Rigby}, {Perrin}, {McElwain}, {Kimble},
  {Friedman}, {Lallo}, {Doyon}, {Feinberg}, {Ferruit}, {Glasse}, {Rieke},
  {Rieke}, {Wright}, {Willott}, {Colon}, {Milam}, {Neff}, {Stark}, {Valenti},
  {Abell}, {Abney}, {Abul-Huda}, {Acton}, {Adams}, {Adler}, {Aguilar}, {Ahmed},
  {Albert}, {Alberts}, {Aldridge}, {Allen}, {Altenburg}, {Alves de Oliveira},
  {Anderson}, {Anderson}, {Anderson}, {Argyriou}, {Armstrong}, {Arribas},
  {Artigau}, {Arvai}, {Atkinson}, {Bacon}, {Bair}, {Banks}, {Barrientes},
  {Barringer}, {Bartosik}, {Bast}, {Baudoz}, {Beatty}, {Bechtold}, {Beck},
  {Bergeron}, {Bergkoetter}, {Bhatawdekar}, {Birkmann}, {Blazek}, {Blome},
  {Boccaletti}, {Boeker}, {Boia}, {Bonaventura}, {Bond}, {Bosley}, {Boucarut},
  {Bourque}, {Bouwman}, {Bower}, {Bowers}, {Boyer}, {Brady}, {Braun}, {Breda},
  {Bresnahan}, {Bright}, {Britt}, {Bromenschenkel}, {Brooks}, {Brooks},
  {Brown}, {Brown}, {Brown}, {Bunker}, {Burger}, {Bushouse}, {Cale}, {Cameron},
  {Cameron}, {Canipe}, {Caplinger}, {Caputo}, {Carey}, {Carniani},
  {Carrasquilla}, {Carruthers}, {Case}, {Chance}, {Chapman}, {Charlot},
  {Charlow}, {Chayer}, {Chen}, {Cherinka}, {Chichester}, {Chilton}, {Chonis},
  {Clark}, {Clark}, {Coe}, {Coleman}, {Comber}, {Comeau}, {Connolly}, {Cooper},
  {Cooper}, {Coppock}, {Correnti}, {Cossou}, {Coulais}, {Coyle}, {Cracraft},
  {Curti}, {Cuturic}, {Davis}, {Davis}, {Dean}, {DeLisa}, {deMeester},
  {Dencheva}, {Dencheva}, {DePasquale}, {Deschenes}, {Hunor Detre}, {Diaz},
  {Dicken}, {DiFelice}, {Dillman}, {Dixon}, {Doggett}, {Donaldson}, {Douglas},
  {DuPrie}, {Dupuis}, {Durning}, {Easmin}, {Eck}, {Edeani}, {Egami},
  {Ehrenwinkler}, {Eisenhamer}, {Eisenhower}, {Elie}, {Elliott}, {Elliott},
  {Ellis}, {Engesser}, {Espinoza}, {Etienne}, {Etxaluze}, {Falini}, {Feeney},
  {Ferry}, {Filippazzo}, {Fincham}, {Fix}, {Flagey}, {Florian}, {Flynn},
  {Fontanella}, {Ford}, {Forshay}, {Fox}, {Franz}, {Fu}, {Fullerton}, {Galkin},
  {Galyer}, {Garcia Marin}, {Gardner}, {Gardner}, {Garland}, {Gasman},
  {Gaspar}, {Gaudreau}, {Gauthier}, {Geers}, {Geithner}, {Gennaro}, {Giardino},
  {Girard}, {Giuliano}, {Glassmire}, {Glauser}, {Glazer}, {Godfrey},
  {Golimowski}, {Gollnitz}, {Gong}, {Gonzaga}, {Gordon}, {Gordon},
  {Goudfrooij}, {Greene}, {Greenhouse}, {Grimaldi}, {Groebner}, {Grundy},
  {Guillard}, {Gutman}, {Ha}, {Haderlein}, {Hagedorn}, {Hainline}, {Haley},
  {Hami}, {Hamilton}, {Hammel}, {Hansen}, {Harkins}, {Harr}, {Hart}, {Hart},
  {Hartig}, {Hashimoto}, {Haskins}, {Hathaway}, {Havey}, {Hayden}, {Hecht},
  {Heller-Boyer}, {Henry}, {Hermann}, {Hernandez}, {Hesman}, {Hicks},
  {Hilbert}, {Hines}, {Hoffman}, {Holfeltz}, {Holler}, {Hoppa}, {Hott},
  {Howard}, {Hunter}, {Hunter}, {Hurst}, {Husemann}, {Hustak}, {Ilinca Ignat},
  {Irish}, {Jackson}, {Jahromi}, {Jakobsen}, {James}, {James}, {Januszewski},
  {Jenkins}, {Jirdeh}, {Johnson}, {Johnson}, {Jones}, {Jones}, {Jones},
  {Jones}, {Jordan}, {Jordan}, {Jurczyk}, {Jurling}, {Kaleida}, {Kalmanson},
  {Kammerer}, {Kang}, {Kao}, {Karakla}, {Kavanagh}, {Kelly}, {Kendrew},
  {Kennedy}, {Kenny}, {Keski-kuha}, {Keyes}, {Kidwell}, {Kinzel}, {Kirk},
  {Kirkpatrick}, {Kirshenblat}, {Klaassen}, {Knapp}, {Knight}, {Knollenberg},
  {Koehler}, {Koekemoer}, {Kovacs}, {Kulp}, {Kumari}, {Kyprianou}, {La Massa},
  {Labador}, {Labiano Ortega}, {Lagage}, {Lajoie}, {Lallo}, {Lam}, {Lamb},
  {Lambros}, {Lampenfield}, {Langston}, {Larson}, {Law}, {Lawrence}, {Lee},
  {Leisenring}, {Lepo}, {Leveille}, {Levenson}, {Levine}, {Levy}, {Lewis},
  {Lewis}, {Libralato}, {Lightsey}, {Link}, {Liu}, {Lo}, {Lockwood}, {Logue},
  {Long}, {Long}, {Loomis}, {Lopez-Caniego}, {Alvarez}, {Love-Pruitt}, {Lucy},
  {Luetzgendorf}, {Maghami}, {Maiolino}, {Major}, {Malla}, {Malumuth},
  {Manjavacas}, {Mannfolk}, {Marrione}, {Marston}, {Martel}, {Maschmann},
  {Masci}, {Masciarelli}, {Maszkiewicz}, {Mather}, {McKenzie}, {McLean},
  {McMaster}, {Melbourne}, {Mel{\'e}ndez}, {Menzel}, {Merz}, {Meyett}, {Meza},
  {Miskey}, {Misselt}, {Moller}, {Morrison}, {Morse}, {Moseley}, {Mosier},
  {Mountain}, {Mueckay}, {Mueller}, {Mullally}, {Murphy}, {Murray}, {Murray},
  {Muzerolle}, {Mycroft}, {Myers}, {Myrick}, {Nanavati}, {Nance}, {Nayak},
  {Naylor}, {Nelan}, {Nickson}, {Nielson}, {Nieto-Santisteban}, {Nikolov},
  {Noriega-Crespo}, {O'Shaughnessy}, {O'Sullivan}, {Ochs}, {Ogle}, {Oleszczuk},
  {Olmsted}, {Osborne}, {Ottens}, {Owens}, {Pacifici}, {Pagan}, {Page},
  {Parrish}, {Patapis}, {Pauly}, {Pavlovsky}, {Pedder}, {Peek},
  {Pena-Guerrero}, {Pennanen}, {Perez}, {Perna}, {Perriello}, {Phillips},
  {Pietraszkiewicz}, {Pinaud}, {Pirzkal}, {Pitman}, {Piwowar}, {Platais},
  {Player}, {Plesha}, {Pollizi}, {Polster}, {Pontoppidan}, {Porterfield},
  {Proffitt}, {Pueyo}, {Pulliam}, {Quirt}, {Quispe Neira}, {Ramos Alarcon},
  {Ramsay}, {Rapp}, {Rapp}, {Rauscher}, {Ravindranath}, {Rawle}, {Regan},
  {Reichard}, {Reis}, {Ressler}, {Rest}, {Reynolds}, {Rhue}, {Richon},
  {Rickman}, {Ridgaway}, {Ritchie}, {Rix}, {Robberto}, {Robinson}, {Robinson},
  {Robinson}, {Rock}, {Rodriguez}, {Rodriguez Del Pino}, {Roellig}, {Rohrbach},
  {Roman}, {Romelfanger}, {Rose}, {Roteliuk}, {Roth}, {Rothwell}, {Rowlands},
  {Roy}, {Royer}, {Royle}, {Rui}, {Rumler}, {Runnels}, {Russ}, {Rustamkulov},
  {Ryden}, {Ryer}, {Sabata}, {Sabatke}, {Sabbi}, {Samuelson}, {Sappington},
  {Sargent}, {Sauer}, {Scheithauer}, {Schlawin}, {Schlitz}, {Schmitz},
  {Schneider}, {Schreiber}, {Schulze}, {Schwab}, {Scott}, {Sembach},
  {Shaughnessy}, {Shaw}, {Shawger}, {Shay}, {Sheehan}, {Shen}, {Sherman},
  {Shiao}, {Shih}, {Shivaei}, {Sienkiewicz}, {Sing}, {Sirianni},
  {Sivaramakrishnan}, {Skipper}, {Sloan}, {Slocum}, {Slowinski}, {Smith},
  {Smith}, {Smith}, {Smith}, {Snyder}, {Soh}, {Sohn}, {Soto}, {Spencer},
  {Stallcup}, {Stansberry}, {Starr}, {Starr}, {Stewart}, {Stiavelli},
  {Straughn}, {Strickland}, {Stys}, {Summers}, {Sun}, {Sunnquist}, {Swade},
  {Swam}, {Swaters}, {Swoish}, {Taylor}, {Taylor}, {Te Plate}, {Tea}, {Teague},
  {Telfer}, {Temim}, {Thatte}, {Thompson}, {Thompson}, {Thomson}, {Tikkanen},
  {Tippet}, {Todd}, {Toolan}, {Tran}, {Trejo}, {Truong}, {Tsukamoto},
  {Tustain}, {Tyra}, {Ubeda}, {Underwood}, {Uzzo}, {Van Campen}, {Vandal},
  {Vandenbussche}, {Vila}, {Volk}, {Wahlgren}, {Waldman}, {Walker}, {Wander},
  {Warfield}, {Warner}, {Wasiak}, {Watkins}, {Weilert}, {Weiser}, {Weiss},
  {Weissman}, {Welty}, {West}, {Wheate}, {Wheatley}, {Wheeler}, {White},
  {Whiteaker}, {Whitehouse}, {Whiteleather}, {Whitman}, {Williams}, {Willmer},
  {Willoughby}, {Wilson}, {Wirth}, {Wislowski}, {Wolf}, {Wolfe}, {Wolff},
  {Workman}, {Wright}, {Wu}, {Wu}, {Wymer}, {Yates}, {Yates}, {Yeager},
  {Yerger}, {Yoon}, {Young}, {Yu}, {Zak}, {Zeidler}, {Zhou}, {Zielinski},
  {Zincke}, \& {Zonak}}]{Rigby2022Characterizatio}
{Rigby}, J., {Perrin}, M., {McElwain}, M., {et~al.} 2022, arXiv e-prints,
  arXiv:2207.05632

\bibitem[{{Robertson} {et~al.}(2013){Robertson}, {Furlanetto}, {Schneider},
  {Charlot}, {Ellis}, {Stark}, {McLure}, {Dunlop}, {Koekemoer}, {Schenker},
  {Ouchi}, {Ono}, {Curtis-Lake}, {Rogers}, {Bowler}, \&
  {Cirasuolo}}]{Robertson2013New-Constraints}
{Robertson}, B.~E., {Furlanetto}, S.~R., {Schneider}, E., {et~al.} 2013, \apj,
  768, 71

\bibitem[{{Sanders} {et~al.}(2020){Sanders}, {Shapley}, {Reddy}, {Kriek},
  {Siana}, {Coil}, {Mobasher}, {Shivaei}, {Freeman}, {Azadi}, {Price}, {Leung},
  {Fetherolf}, {de Groot}, {Zick}, {Fornasini}, \&
  {Barro}}]{Sanders2020The-MOSDEF-surv}
{Sanders}, R.~L., {Shapley}, A.~E., {Reddy}, N.~A., {et~al.} 2020, \mnras, 491,
  1427

\bibitem[{{Stefanon} {et~al.}(2022){Stefanon}, {Bouwens}, {Illingworth},
  {Labb{\'e}}, {Oesch}, \& {Gonzalez}}]{Stefanon2022High-Equivalent}
{Stefanon}, M., {Bouwens}, R.~J., {Illingworth}, G.~D., {et~al.} 2022, \apj,
  935, 94

\bibitem[{{Tang} {et~al.}(2019){Tang}, {Stark}, {Chevallard}, \&
  {Charlot}}]{Tang2019MMT/MMIRS-spect}
{Tang}, M., {Stark}, D.~P., {Chevallard}, J., \& {Charlot}, S. 2019, \mnras,
  489, 2572

\bibitem[{{Tang} {et~al.}(2022){Tang}, {Stark}, \&
  {Ellis}}]{Tang2022Stellar-populat}
{Tang}, M., {Stark}, D.~P., \& {Ellis}, R.~S. 2022, \mnras, 513, 5211

\bibitem[{{Taylor} {et~al.}(2022){Taylor}, {Barger}, \&
  {Cowie}}]{Taylor2022Metallicities-o}
{Taylor}, A.~J., {Barger}, A.~J., \& {Cowie}, L.~L. 2022, arXiv e-prints,
  arXiv:2208.06418

\bibitem[{{Trump} {et~al.}(2022){Trump}, {Arrabal Haro}, {Simons}, {Backhaus},
  {Amor{\'\i}n}, {Dickinson}, {Fern{\'a}ndez}, {Papovich}, {Nicholls},
  {Kewley}, {Brunker}, {Salzer}, {Wilkins}, {Almaini}, {Bagley}, {Berg},
  {Bhatawdekar}, {Bisigello}, {Buat}, {Burgarella}, {Calabr{\`o}}, {Casey},
  {Ciesla}, {Cleri}, {Cole}, {Cooper}, {Cooray}, {Costantin}, {Ferguson},
  {Finkelstein}, {Fujimoto}, {Gardner}, {Gawiser}, {Giavalisco}, {Grazian},
  {Grogin}, {Hathi}, {Hirschmann}, {Holwerda}, {Huertas-Company}, {Hutchison},
  {Jogee}, {Juneau}, {Jung}, {Kartaltepe}, {Kirkpatrick}, {Koekemoer}, {Lotz},
  {Lucas}, {Magnelli}, {Matharu}, {P{\'e}rez-Gonz{\'a}lez}, {Pirzkal},
  {Rafelski}, {Rose}, {Seill{\'e}}, {Somerville}, {Straughn}, {Tacchella},
  {Vanderhoof}, {Weiner}, {Wuyts}, {Yung}, \&
  {Zavala}}]{Trump2022The-Physical-Co}
{Trump}, J.~R., {Arrabal Haro}, P., {Simons}, R.~C., {et~al.} 2022, arXiv
  e-prints, arXiv:2207.12388

\bibitem[{{Trussler} {et~al.}(2022){Trussler}, {Adams}, {Conselice},
  {Ferreira}, {Austin}, {Bhatawdekar}, {Caruana}, {Lovell}, {Roper}, {Verma},
  {Vijayan}, \& {Wilkins}}]{Trussler2022Seeing-sharper-}
{Trussler}, J. A.~A., {Adams}, N.~J., {Conselice}, C.~J., {et~al.} 2022, arXiv
  e-prints, arXiv:2207.14265

\end{thebibliography}

\end{document}